\documentclass[aps,prl,reprint,twocolumn,showpacs,superscriptaddress]{revtex4-1}

\usepackage{graphicx,graphics,amsfonts,amsmath,amssymb}

\usepackage{color}
\usepackage{verbatim}

\newcommand{\beginsupplement}{%
        \setcounter{table}{0}
        \renewcommand{\thetable}{S\arabic{table}}%
        \setcounter{figure}{0}
        \renewcommand{\thefigure}{S\arabic{figure}}%
        }

\begin{document}

\title{Critical Vortex Shedding in a Strongly Interacting Fermionic Superfluid}

\author{Jee Woo Park}
\affiliation{Department of Physics and Astronomy, and Institute of Applied Physics, Seoul National University, Seoul 08826, Korea}
\author{Bumsuk Ko}
\affiliation{Department of Physics and Astronomy, and Institute of Applied Physics, Seoul National University, Seoul 08826, Korea}
\affiliation{Center for Correlated Electron Systems, Institute for Basic Science, Seoul 08826, Korea}
\author{Y. Shin}
\email{yishin@snu.ac.kr}
\affiliation{Department of Physics and Astronomy, and Institute of Applied Physics, Seoul National University, Seoul 08826, Korea}
\affiliation{Center for Correlated Electron Systems, Institute for Basic Science, Seoul 08826, Korea}

\begin{abstract}
We study the critical vortex shedding in a strongly interacting fermionic superfluid of $^{6}$Li across the BEC-BCS crossover. By moving an optical obstacle in the sample and directly imaging the vortices after time of flight, the critical velocity $v_{{\rm c}}$ for vortex shedding is measured as a function of the obstacle travel distance $L$. The observed $v_{\rm c}$ increases with decreasing $L$, where the rate of increase is the highest in the unitary regime. In the deep BEC regime, an empirical dissipation model well captures the dependence of $v_{{\rm c}}$ on $L$, characterized by a constant value of $\eta = -\frac{{\rm d}(1/v_{{\rm c}})}{{\rm d}(1/L)}$. However, as the system is tuned across the resonance, a step increase of $\eta$ develops about a characteristic distance $L_{\rm c}$ as $L$ is increased, where $L_{\rm c}$ is comparable to the obstacle size. This bimodal behavior is strengthened as the system is tuned towards the BCS regime. We attribute this evolution of $v_{{\rm c}}$ to the participation of pair breaking in the vortex shedding dynamics of a fermionic superfluid. 
\end{abstract}

\maketitle

Superfluidity, the absence of friction in particle flow, is a spectacular demonstration of macroscopic quantum coherence. One of its defining properties is the existence of a critical velocity, above which the creation of fundamental excitations gives rise to drag and dissipation in the superfluid. From a microscopic perspective, the Landau criterion presents a critical velocity $v_{\rm L} = \textrm{min}_{p}[\epsilon(p)/p]$, where $\epsilon(p)$ is the energy of the superfluid's microscopic excitation with momentum $p$. However, it is known that when a superfluid flows past an obstacle larger than the healing length $\xi$, the nucleation of quantized vortices lowers the critical velocity below $v_{\rm L}$, as first observed in superfluid $^{4}$He~\cite{Donnelly1991}, and that their creation strongly modifies the thermodynamic and transport response of the superfluid system~\cite{Blatter1994, Vinen2002}. Despite their importance, the details of the vortex nucleation process and its relation to the superfluid's microscopic modes of excitation remain as open questions, in particular, in the study of strongly correlated quantum fluids~\cite{Anderson1966, Bulgac2011a, Varoquaux2015}.  

Strongly interacting atomic Fermi gases with tunable interactions offer an interesting opportunity to investigate the microscopic aspects of superfluid dissipation by accessing the crossover between Bose-Einstein condensation (BEC) and Bardeen-Cooper-Schrieffer (BCS) superfluidity~\cite{grei03molbec, joch03bec, zwie03molBEC, rega04, zwie04rescond, bour04coll, chin04gap, Zwierlein2005}. In the crossover, the nature of superfluidity smoothly changes from bosonic to fermionic, and the elementary excitation determining the Landau criterion is transformed from phonons to fermionic quasiparticles via pair breaking~\cite{Combescot2006}. Superfluid dissipation in the BEC-BCS crossover has been explored in a number of experiments, where the critical velocity for heating in the presence of a moving obstacle or in a counterflow of bosonic and fermionic superfluids was measured~\cite{Miller2007, Weimer2015, Delehaye2015}, and recently, the emergence of dissipation in a Josephson junction by phase slip and vortex nucleation was studied~\cite{Valtolina2015, Burchianti2018a}. However, the roles of the coexisting elementary excitations and their possible interplay in the dissipative nucleation of quantized vortices remain obscure.

In this Letter, we report on the measurement of the critical velocity $v_{\rm c}$ for vortex shedding in a strongly interacting fermionic superfluid of $^{6}$Li across the BEC-BCS crossover. The response of the superfluid to a pulsed linear motion of an obstacle over a finite distance $L$ results in a characteristic dependence of $v_{\rm c}$ on $L$. A general tendency for $v_{\rm c}$ to increase with decreasing $L$, which is predictable from energy considerations, is observed for all the investigated interaction strengths. The maximal rate of increase is observed near unitarity, implying that Fermi superfluidity is the most robust in this regime. In the BEC limit, we find that the characteristic relationship between $v_{\rm c}$ and $L$ is described by a constant value of $\eta = -\frac{{\rm d}(1/v_{{\rm c}})}{{\rm d}(1/L)}$, which is in accordance with an empirical model that assumes a linear dependence of the dissipation rate on the obstacle velocity. In the crossover regime, however, we observe a sudden jump of $\eta$ at a characteristic distance $L_{\rm c}$ as $L$ is increased, where $L_{\rm c}$ is comparable to the obstacle size, and this jump becomes larger when approaching the BCS regime. We argue that this evolution is attributable to the emergence of pair breaking in a fermionic superfluid. Our results shed light on the role of microscopic excitations in vortex nucleation and also provide a stringent benchmark for time-dependent theories of strongly interacting fermionic systems~\cite{Bulgac2011, Ancilotto2013a, Forbes2014}.


\begin{figure}
\includegraphics[width=8.5cm]{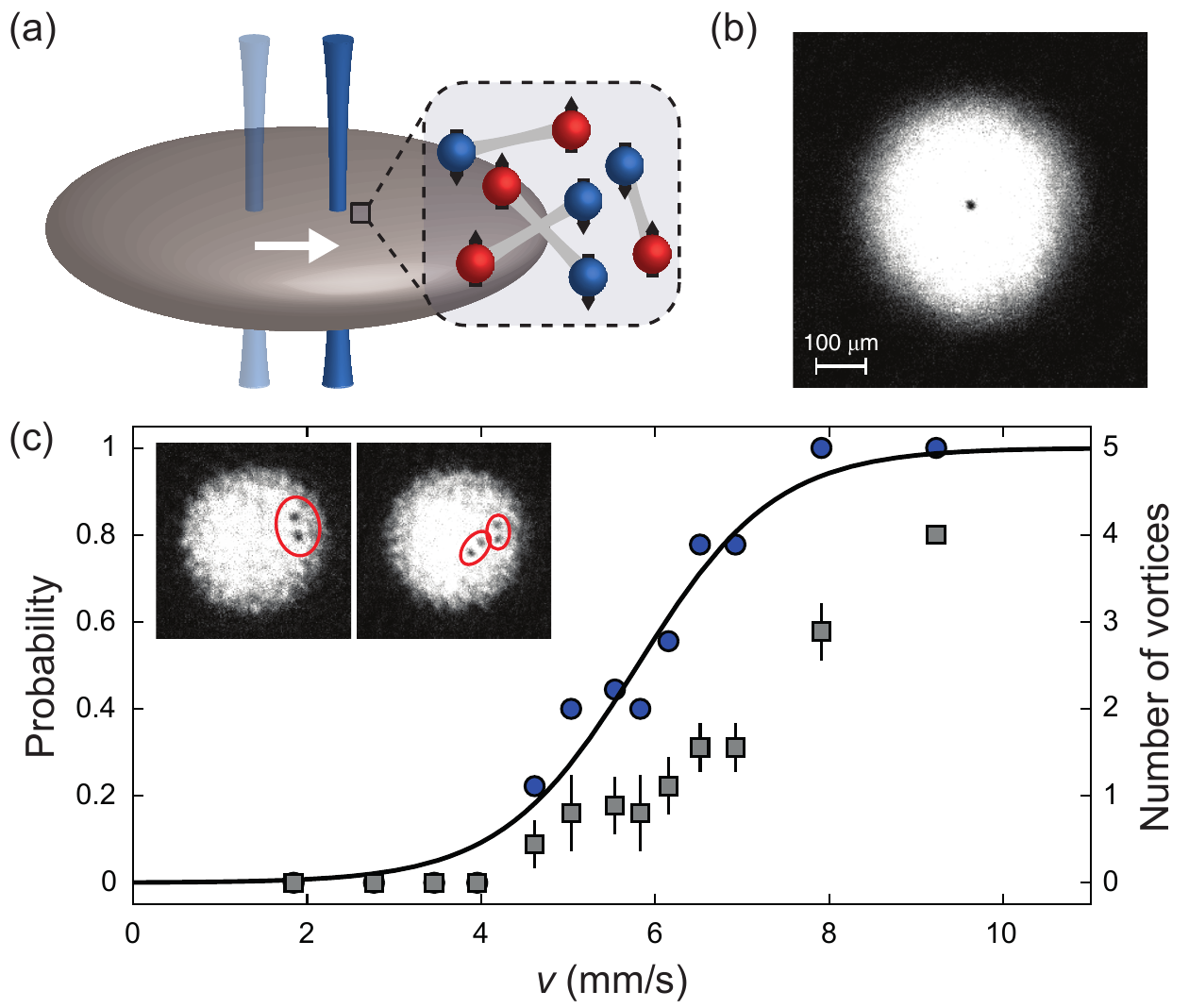}
\caption{Vortex shedding in a strongly interacting fermionic superfluid. (a) Schematic of the experiment. A cylindrical impenetrable obstacle, consisting of a focused repulsive Gaussian laser beam, is translated at a constant velocity about the center of a disc-shaped strongly interacting fermionic condensate of $^{6}$Li. (b) An \textit{in situ} image of the condensate at unitarity penetrated by the obstacle. (c) The number of vortices (gray squares) and the probability of observing vortex dipoles (blue circles) after sweeping the obstacle as a function of $v$. Each data point comprises at least 9 realizations of the same experiment. The black solid line is a sigmoidal fit to the probability. }
\label{Fig1}
\end{figure}

The starting point of the experiment is the creation of a strongly interacting fermionic superfluid of $^{6}$Li in an optical dipole trap. To this end, the experimental apparatus described in Ref.~\cite{Heo2011} has been modified to accommodate $^{6}$Li together with bosonic $^{23}$Na. In the experiment, both species are simultaneously loaded into a magneto-optical trap, and then optically pumped and transferred to a plugged magnetic quadrupole trap, where forced radio-frequency (rf) evaporation of $^{23}$Na sympathetically cools $^{6}$Li to quantum degeneracy~\cite{Hadzibabic2003,vortex_supp}. The resulting $^{6}$Li atoms are then loaded into an optical dipole trap formed by focusing a 1064 nm laser beam.

To access the strongly interacting regime, we use a broad $s$-wave Feshbach resonance between the two lowest hyperfine states of $^{6}$Li (denoted $|{1}\rangle$ and $|{2}\rangle$) located at 832 G that allows precise tuning of the $s$-wave scattering length $a$~\cite{Zurn2013}. Initially, all the atoms are transferred to $|{1}\rangle$, and then an equal mixture of the two states is prepared near 870~G by using Landau-Zener rf-sweeps. The final stage of evaporation is performed at 815 G by reducing the dipole trap laser intensity. After evaporation, the magnetic field is adiabatically ramped to a value where the critical velocity measurement will be performed. At unitarity, this procedure produces a superfluid sample consisting of $N\approx 1.0 \times 10^{6}$ $^{6}$Li atoms per spin state with a typical condensate fraction of approximately 80$\%$, corresponding to a temperature of ${T/T_{\rm F}<0.1}$~\cite{Chen2005a}. Here, $T_{\rm F}$ is the Fermi temperature defined as $T_{\rm F}{=}E_{\rm F}/k_{\rm B} $, where $E_{\rm F}{=}\hbar^2k_{\rm F}^2/2m{=}\hbar \overline{\omega} (6N)^{1/3}$ is the Fermi energy of a non-interacting Fermi gas, $\hbar$ is the reduced Planck constant, $k_{\rm F}$ is the Fermi wavenumber, $m$ is the atomic mass of $^{6}$Li, and $\overline{\omega}$ is the geometric mean of the trap frequencies. The final trapping frequencies are $(\omega_{x}, \omega_{y}, \omega_{z}) =$ $2 \pi \times (17, 18, 483)$ Hz, where the radially symmetric confinement is mainly provided by the residual magnetic field curvature from the Feshbach field, and the tight $z$-confinement is provided by the optical dipole trap. Such an oblate geometry is favored for studies of quantum vortices, since it provides a preferred direction for a vortex line that is stable against bending excitations~\cite{Neely2010, Kwon2015}.

\begin{figure}
\includegraphics[width=8.5cm]{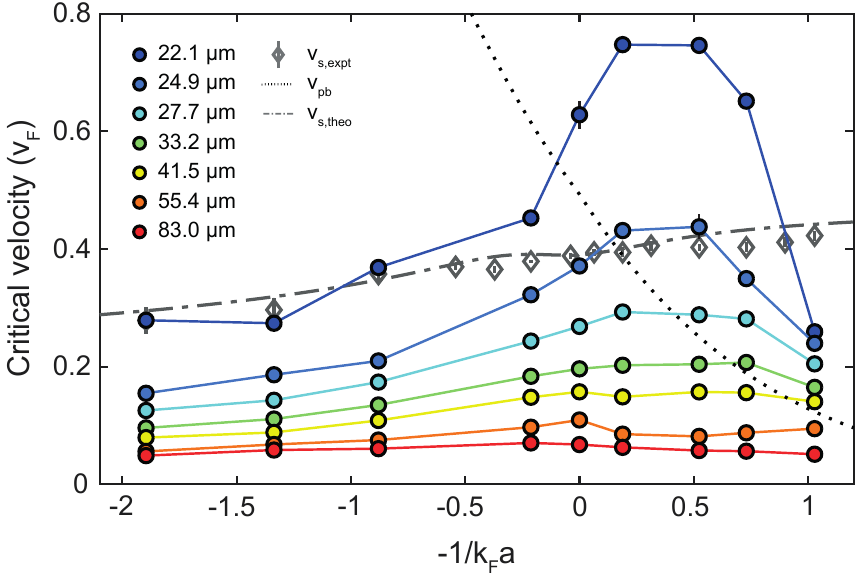}
\caption{The measured critical velocity for vortex shedding $v_{\rm c}$ (filled circles) at different sweeping lengths $L$ and the speed of sound $v_{\rm s, exp}$ (open diamonds) in the BEC-BCS crossover, in units of the Fermi velocity $v_{\rm F}$. The error bars are one standard deviation of the fit. The gray dot-dashed line is the theoretical speed of sound $v_{\rm s,theo}$ from  quantum Monte Carlo calculations, for column-averaged densities~\cite{Manini2005}. The black dotted curve is the pair breaking velocity $v_{\rm pb}$ from the the mean field BCS theory.}
\label{Fig2}
\end{figure}

\begin{figure*}
\includegraphics[width=18cm]{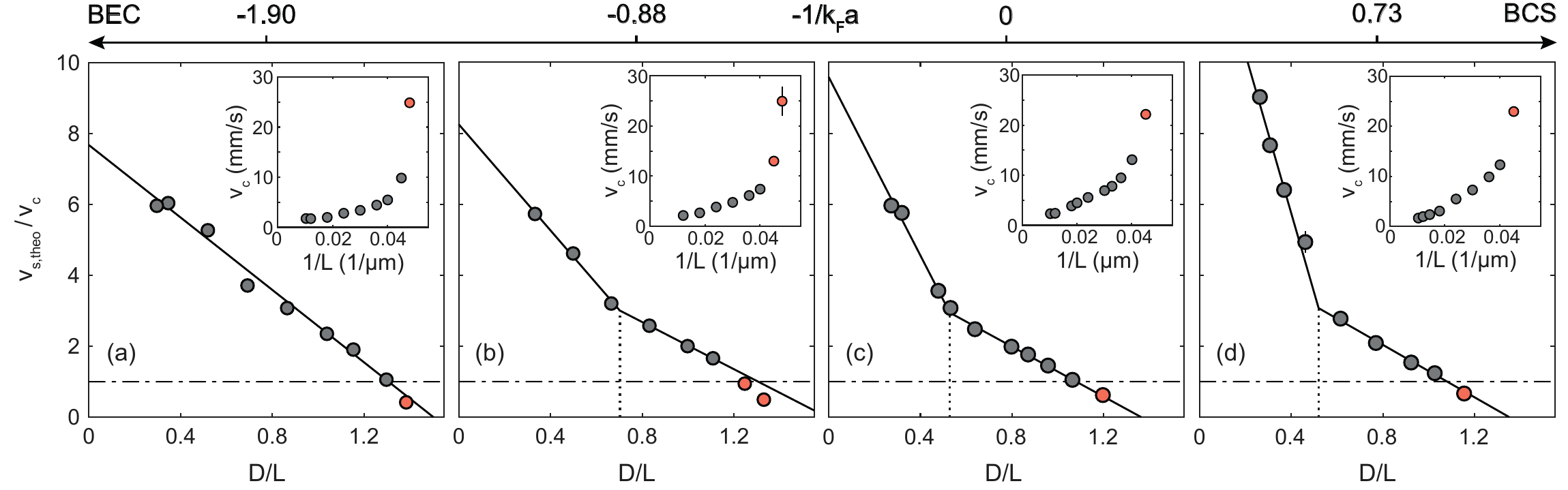}
\caption{Inverse critical velocity $1/v_{\rm c}$ as a function of the inverse sweeping distance $1/L$. $v_{\rm c}$ and $L$ are nomalized by the speed of sound $v_{\rm s, theo}$ and the effective obstacle diameter $D$, respectively. Four representative graphs whose $-1/k_{\rm F}a$ is equal to (a) $-1.9$, (b) $-0.88$, (c) 0, and (d) 0.73 are shown. The solid line is the fit of the dissipation model to the data points with $v_{\rm c} < v_{\rm s,theo}$ (gray) while excluding those with $v_{\rm c} > v_{\rm s, theo}$ (red). The dotted line indicates the characteristic $D/L_{\rm c}$ where the bimodality develops. The dot-dashed line marks where $v_{\rm c} = v_{\rm s,theo}$. The error bars are one standard deviation of the fit. The insets show the values of $v_{\rm c}$ as a function of $1/L$.}
\label{Fig3}
\end{figure*}

A schematic of the experiment is shown in Fig~\ref{Fig1}(a). A repulsive optical obstacle is translated about the center of the sample by a fixed distance $L$ during a variable time $t$ at a constant velocity $v=L/t$. After the sweep, the Feshbach field is switched off and simultaneously the sample is released from the trap for time-of-flight expansion, during which the vortices expand in the radial direction relative to the condensate. Following the expansion, an absorption image of the condensate is taken at $B=690$ G. A representative set of images displaying one and two generated vortex dipoles are shown in the inset of Fig.~\ref{Fig1}(c). The critical velocity for vortex shedding is extracted from a sigmoidal fit to the probability $P(v)$ of observing vortex-antivortex pairs as $P(v)~{=}~1/(1+e^{-(v-v_{\rm c})/\sigma})$, where $P(v)$ is obtained by varying the time $t$ for a given $L$ [Fig.~\ref{Fig1}(c)].

The repulsive optical obstacle consists of a tightly focused 532 nm Gaussian laser beam propagating along the $z$-axis, whose position is controlled using a piezo-actuated mirror~\cite{Kwon2015}. The $1/e^{2}$ radius of the beam is measured to be $w_{0}=9.5$ $\mu$m, which is an order of magnitude larger than the Fermi length scale $1/k_{\rm F} \approx 0.30~\mu$m. The height of the obstacle is set at $k_{B} \times 10$ $\mu$K, which is about 20 times higher than the Fermi energy $E_{\rm F}\approx k_{B} \times 0.45$ $\mu$K. Comparing the obstacle height to the chemical potential of the sample, the obstacle has an effective diameter of $D \approx 26 \mu$m at unitarity~\cite{vortex_supp, Kwon2015}. The radial Thomas-Fermi radius $R_{\mathrm{TF}, r}$ of the sample at unitarity is $ 260 ~\mu$m, which is significantly larger than the obstacle sweeping distance $L$ ranging from 20 to 90 $\mu$m. $R_{\mathrm{TF}, z}$ in the z-direction, $ 12 ~\mu$m, is much shorter than the Rayleigh length of the obstacle beam, $ 590 ~\mu$m, ensuring that its divergence is negligible within the condensate.

When employing the obstacle beam, precaution is taken to avoid exciting unwanted dynamics in the condensate. The beam is switched on at the starting point of the sweep in 100 ms, and a hold time of 50 ms is applied. After the sweep, an extra hold time of 50 ms is provided, and then the beam is linearly switched off in 200 ms. The effect of the ramp and the hold times on the measured critical velocities is studied, and their values are carefully chosen to ensure that they are adiabatic and independent of the measurements~\cite{vortex_supp}.

The vortex shedding critical velocities for various sweeping distances are measured for a broad range of interaction strengths covering the full BEC-BCS crossover, as shown in Fig.~\ref{Fig2}. As a reference, the speed of sound $v_{\rm s,exp}$ of the superfluid is also measured from the propagating speed of an outgoing circular density wave, which is triggered by abruptly switching off the obstacle beam depleting the center of the sample~\cite{vortex_supp, Joseph2007, Weimer2015}. This is shown together with a theoretical speed of sound $v_{\rm s, theo}$ for the column averaged density from quantum Monte Carlo calculations~\cite{Astrakharchik2004, Manini2005} and the mean-field BCS pair breaking velocity $v_{\rm pb}$. $v_{\rm s,exp}$ manifests the speed of sound of the column averaged density since our sample is hydrodynamic in the $z$-axis. Here, the theory curves are scaled to our definition of $k_{\rm F}$ using the equation of state while assuming the local density approximation.

The most notable feature in Fig.~\ref{Fig2} is the dramatic increase of $v_{\rm c}$ near unitarity as the sweeping distance $L$ is reduced. The rise of $v_{\rm c}$ with decreasing $L$ is expected in general since the necessary energy for vortex nucleation has to be provided within a shorter obstacle travel distance. The enhanced rate of increase near unitarity reflects the stability of the Fermi superfluid against vortex nucleation in the strongly interacting regime, which is consistent with previous experiments that reported the maximal robustness of fermionic superfluidity near unitarity~\cite{Miller2007,Weimer2015, Valtolina2015, Burchianti2018a}. However, a comparison between $v_{\rm c}$ and $v_{\rm L}$ has to be made prudently since $v_{\rm c}$ probes the vortex nucleation dynamics within a finite distance $L$, whereas $v_{\rm L}$ signifies the onset of dissipation via the creation of microscopic excitations. One may conjecture that $v_{\rm c}$ will converge to $v_{\rm L}$ for infinitely long $L$, where a scaling factor determined by the obstacle geometry may be involved.

Another surprise comes from the observation of well-defined vortex shedding critical velocities exceeding the speed of sound for sufficiently short $L$. When an object moves through a medium faster than its speed of sound, shock waves are created that lead to strong density modulations. Furthermore, the ensuing flow will be turbulent, and vortex shedding will be highly irregular~\cite{Ancilotto2013a, wini00drag}. Thus, it is out of expectation that no abrupt change in the response of $v_{\rm c}$ is observed when the obstacle is translated faster than the speed of sound over a short distance. In fact, this behavior is observed throughout the explored crossover regimes when $L$ is sufficiently reduced (red data points in Fig.~\ref{Fig3}), suggesting that it may be a general characteristic of a compressible gaseous superfluid. In a numerical simulation of vortex shedding in a unitary Fermi superfluid, strong density accumulation in front of a stirring obstacle has been observed~\cite{Bulgac2011}. This may increase the local critical velocity, allowing superflow above the nominal critical velocity.

\begin{figure}
\includegraphics[width=8.5cm]{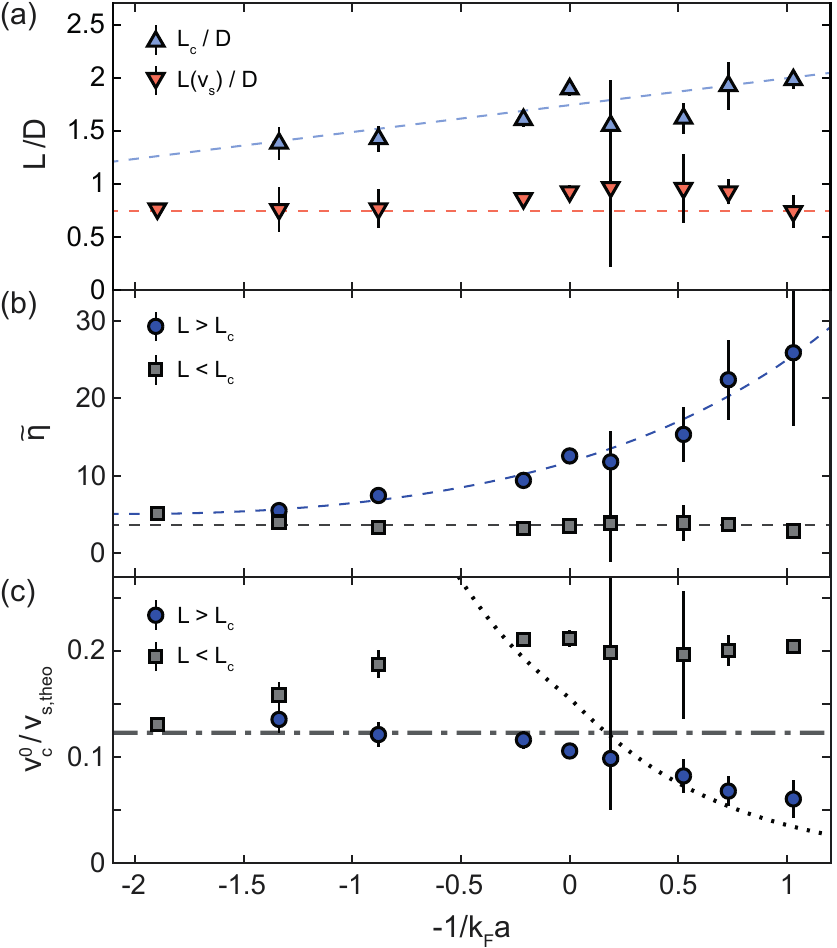}
\caption{Characterization of the evolution of $v_{\rm c}(L)$ in the BEC-BCS crossover. (a) The sweeping distance $L_{\rm c}$ at which a step jump of $\widetilde\eta$, the magnitude of the slope of the fit, is observed (blue triangles), and the value of $L$ where $v_{\rm c}=v_{\rm s,theo}$ (inverted red triangles). The dashed lines are guide to the eye. (b) $\widetilde\eta$ for $L>L_{\rm c}$ (blue circles) and $L<L_{\rm c}$ (gray squares). The blue and gray dashed lines are a guide to the eye. (c) The inverse of the $y$-intercept of the fit, equal to the critical velocity for drag normalized by the speed of sound, for $L>L_{\rm c}$ (blue circles) and $L<L_{\rm c}$ (gray squares). The speed of sound (dot-dashed line) and the pair breaking velocity (dotted line) normalized by the speed of sound with a multiplicative factor are shown together. The error bars are one standard deviation of the fit.}
\label{Fig4}
\end{figure}

To further elucidate the dependence of $v_{\rm c}$ on $L$, we adopt a simple dissipation model where an obstacle moving faster than the critical velocity $v_{\rm c}^{0}$ for the appearance of a drag force deposits energy into the system at a rate given by $P=\Gamma (v-v_{\rm c}^{0})$~\cite{Frisch1992, Winiecki1999, Kwon2015}. Here, $\Gamma$ is a proportionality constant that captures the energy transfer efficiency with dimensions of force. When the energy cost of exciting a vortex dipole is $E_{\rm c}$, we have $E_{\rm c}=Pt=\Gamma (v_{\rm c}-v_{\rm c}^0) (L/v_{\rm c})$, which can be re-expressed as
\begin{align}
v_{\rm c}(L)=\frac{v_{\rm c}^0}{1-l_{0}/L},
\label{Eq1}
\end{align}
where $l_{0}{=} E_{\rm c}/\Gamma$. Note that $l_{0}$ denotes the minimum distance required to shed a vortex dipole. This model motivates us to plot the inverse of $v_{\rm c}$ versus $1/L$ for each interaction strength, as shown in Fig.~\ref{Fig3}. Here, $v_{\rm c}$ and $L$ are normalized with respect to the speed of sound $v_{\rm s, theo}$ and the obstacle diameter $D$ at each $1/k_{\rm F}a$, respectively. We find that the model fits the data exceptionally well in the far BEC regime [Fig.~\ref{Fig3}(a)], where the change of $v_{\rm c}$ with $L$ is characterized by a constant magnitude of the slope, ${\widetilde\eta}=\eta \times v_{\rm s, theo}/D$.  We note that our measurements indicate $v_{\rm c}^{\rm 0} = 0.13~v_{\rm s,theo}$, which is about three times lower than the previous result obtained for a weakly interacting BEC with a highly oblate geometry~\cite{Kwon2015, Kwon2015a}. This discrepancy may arise from the stronger three-dimensional nature of our sample~\cite{Jackson2000b, wini00drag}.

Approaching the resonance, however, the experimental data deviates from the model, showing a sudden jump of $\widetilde\eta$ near $L_{\rm c}\sim 1.5~D$ as $L$ is increased. This bimodal structure emerges already on the BEC side near the Feshbach resonance and becomes further pronounced as the interaction is tuned towards the BCS regime. We characterize this evolution by applying a bilinear fit to the data while excluding the data points with $v_{\rm c}>v_{\rm s, theo}$, as shown in Fig.~\ref{Fig3}(b)-(d). The fitted values of $L_{\rm c}/D$ exhibit a slightly increasing trend towards the BCS limit [Fig.~\ref{Fig4}(a)], and the ratio of $\widetilde\eta$ in the long $L>L_{\rm c}$ branch to that in the short $L<L_{\rm c}$ branch increases by up to almost an order of magnitude as $1/k_{\rm F}a$ \hbox{approaches $-1$ [Fig.~\ref{Fig4}(b)].}

The $y$-intercept of the fit represents the inverse of the critical velocity for the appearance of drag, $v_{\rm c}^{0}$, at the investigated interaction strength, and we compare the obtained $v_{\rm c}^{0}$ with the Landau critical velocity $v_{\rm L} = \textrm{min}(v_{\rm s},v_{\rm pb})$~\cite{Combescot2006, Miller2007, Weimer2015}. We find that the overall trend of $v_{\rm c}^{0}$ closely follows $v_{\rm L}$ in the BEC-BCS crossover, as shown in Fig.~\ref{Fig4}(c), which suggests that the onset of the drag force leading to vortex dipole emission originates from the microscopic excitations of the superfluid. Specifically, the suppression of $v_{\rm c}^{0}/v_{\rm s, theo}$ towards the BCS regime reveals the participation of the pair breaking mechanism in the vortex shedding dynamics across the crossover.

It is remarkable that $\widetilde\eta$ for short $L<L_{\rm c}$ is nearly constant at the value established in the far BEC regime throughout the crossover [Fig.~\ref{Fig4}(b)]. Since pair breaking effects must play a negligible role in the BEC regime, this observation may suggest that the role of pair breaking excitations in the vortex nucleation dynamics is suppressed in the short $L$ branch, even away from the BEC limit. Looking closely, however, if we linearly extrapolate the data with $L<L_{\rm c}$ and extract the critical velocity $v_{\rm c}^{0}(L<L_{\rm c})$, its value increases with respect to the speed of sound until it reaches a plateau around the resonance [Fig.~\ref{Fig4}(c)], showing that the vortex nucleation for short $L < L_{\rm c}$ cannot be explained solely by the speed of sound.

Our measurements reveal the involvement of pair breaking excitations in vortex shedding in a strongly interacting fermionic superfluid, but they pose a number of questions as to what determines the critical distance $L_{\rm c}$ and why the transition across $L_{\rm c}$ is so sharp. A further inquiry is necessary to understand the dynamical interplay between the bosonic and fermionic elementary excitations in the vortex nucleation process.

In conclusion, we have measured the critical velocity for vortex shedding and studied its change according to the obstacle sweeping distance in a strongly interacting fermionic condensate across the BEC-BCS crossover. A steep increase of the critical velocity is observed near unitarity, demonstrating the robustness of Fermi superfluidity in this regime, and the participation of pair breaking excitations in the vortex shedding mechanism is revealed from the change. In light of the recent experimental observation of a von Karman vortex street in a weakly interacting BEC~\cite{Kwon2016}, it would be intriguing to investigate the universality of the vortex shedding behavior in the current system, e.g., in terms of $L_{\rm c}/D$ and $l_0/D$.

We thank Aurel Bulgac for helpful discussions. This work was supported by the Institute for Basic Science in Korea (Grant No. IBS-R009-D1) and the National Research Foundation of Korea (Grant No. NRF-2018R1A2B3003373). JWP acknowledges support from the POSCO Science Fellowship of the POSCO TJ Park Foundation.

\bibliographystyle{apsrev4-1}

\clearpage
\beginsupplement

\section{Supplemental Material}

\subsection{Sample preparation}

To prepare a strongly interacting Fermi gas of $^{6}$Li, we first load fermionic $^{6}$Li and bosonic $^{23}$Na atoms simultaneously into a magneto-optical trap (MOT) from a Na-Li oven. Then, the atoms are optically pumped to their magnetically trappable stretched hyperfine states $^{6}$Li$\vert F{=}3/2, m_{F}{=}3/2 \rangle$ and $^{23}$Na$\vert F{=}2, m_{F}{=}2 \rangle$, respectively, and subsequently transferred to an optically plugged magnetic trap. The optical plug consists of a tightly focused 532 nm Gaussian laser beam with a $1/e^2$ radius of 45 $\mu$m. In the magnetic trap, the mixture is cooled by performing radio frequency (rf) evaporation of $^{23}$Na on the $\vert F{=}2, m_{F}{=}2 \rangle \rightarrow \vert F{=}1, m_{F}{=}1 \rangle$ transition, while $^{6}$Li is sympathetically cooled via collisions with $^{23}$Na. Then, the sample is transferred to a single beam optical dipole trap (ODT) that features an aspect ratio of 110:1, and a resonant light pulse is applied to selectively remove the residual $^{23}$Na atoms. 

To access the $^{6}$Li Feshbach resonance between $|1\rangle=\vert F{=}1/2, m_{F}{=}1/2 \rangle$ and $|2\rangle=\vert F{=}1/2, m_{F}{=}-1/2 \rangle$ at 832 G, the atoms are initially transferred to $|1\rangle$ by applying a rf Landau-Zener sweep at a magnetic field of 3 G, and then the field is increased to 870G in 500 ms, where an additional Landau-Zener sweep prepares the atoms in an equal mixture of $|1\rangle $ and $|2\rangle$. A magnetic field ramp to 815 G in 100 ms converts the atom pairs into Feshbach molecules, and the final stage of evaporation is performed at this field by lowering the ODT trap depth. The magnetic field is parallel to the tightly confining axis of the ODT. At the end of the evaporation, the ODT trap depth is increased by a factor of 1.5 to ensure that the final trap depth is deeper than the Fermi energy $E_{\rm F}$ of the sample such that no atom loss occurs when accessing the BCS side of the resonance. Finally, the magnetic field is adiabatically ramped to a value where the critical velocity measurement will be performed. At resonance, the final trapping frequencies are $(\omega_{x}, \omega_{y}, \omega_{z})$ = $2 \pi \times (17.2, 18.0, 483)$ Hz, where the radially symmetric confinement mainly comes from the residual magnetic curvature of the Feshbach field while the tight axial trapping comes from the ODT. The maximal variation of the radial trapping frequencies for the investigated Feshbach fields is less than 10$\%$. The typical atom number per spin state is about $1 \times 10^{6}$, corresponding to  $E_{F} = \hbar \bar{\omega} (6N)^{1/3} \sim 0.5 ~\mu $K, where $\bar{\omega}$ is the geometric mean of the trap frequencies.

\begin{figure}[t]
\includegraphics[width=8.5cm]{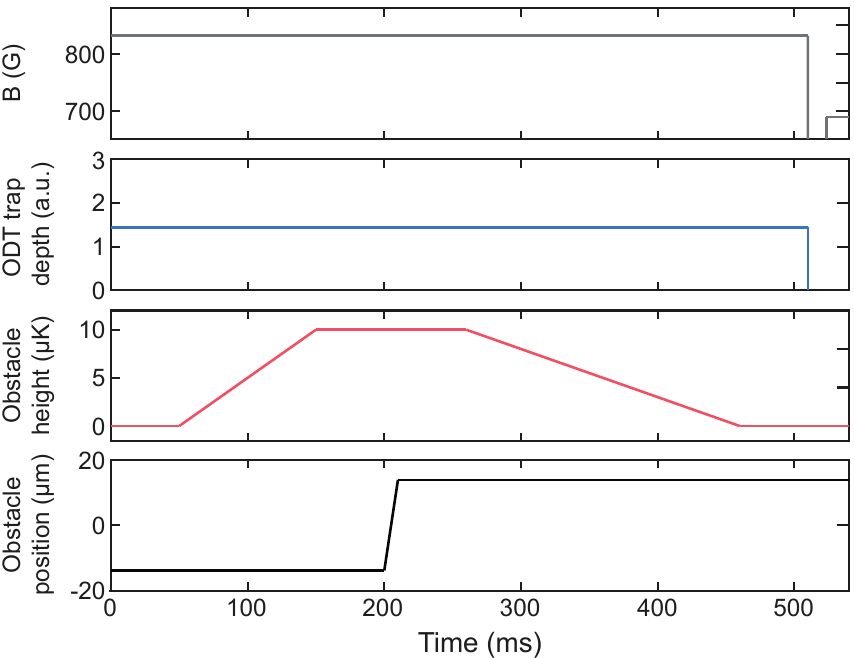}
\caption{An example of the experimental sequence after the magnetic field is set at the desired value. The ramp and hold times of the obstacle height are carefully chosen to be independent of the measurement of the critical velocity.}
\label{S1}
\end{figure}

\subsection{Obstacle translation}
The critical velocity for vortex shedding is investigated by translating about the center of the condensate an impenetrable cylindrical obstacle formed by focusing a 532 nm repulsive Gaussian laser beam. The experimental sequence during the measurement is shown in Fig.~\ref{S1}. A number of precautions were taken to avoid unwanted dynamics in the superfluid during the investigation. First, when the final ODT trap depth is reached, the sample is held for 0.5 s to ensure that it reaches the equilibrium. Then, the magnetic field is adiabatically ramped in 300 ms to the desired value where the investigation takes place. Since the potential minimum of the ODT and the magnetic field curvature do not perfectly coincide, it is crucial to ramp the magnetic field slowly enough throughout the whole experimental sequence to suppress undesirable excitations such as dipole mode oscillations. At the end of the ramp, we apply an additional hold time of 50 ms. Subsequently, the optical obstacle is switched on adiabatically in 100 ms, followed by a 50 ms hold time, to ensure that the measurement is not affected by the switch-on process. The obstacle is swept by a given distance about the center of the sample using a piezo-driven mirror. For a chosen travel distance $L$, the velocity of the obstacle is controlled by the duration of the translation. After completing the sweep, the obstacle stays still for 50 ms, and then its intensity is linearly ramped down to zero in 200 ms. The effect of the ramp down time on the measured critical velocity is studied in detail and is discussed in the following section. From our studies, 200 ms of ramp down time is found to be sufficiently long for the measurement to be independent. Finally, after an additional 50 ms of hold time, the Feshbach field is switched off, and simultaneously the sample is released from the trap for time-of-flight imaging.

\begin{figure}[t]
\includegraphics[width=8.5cm]{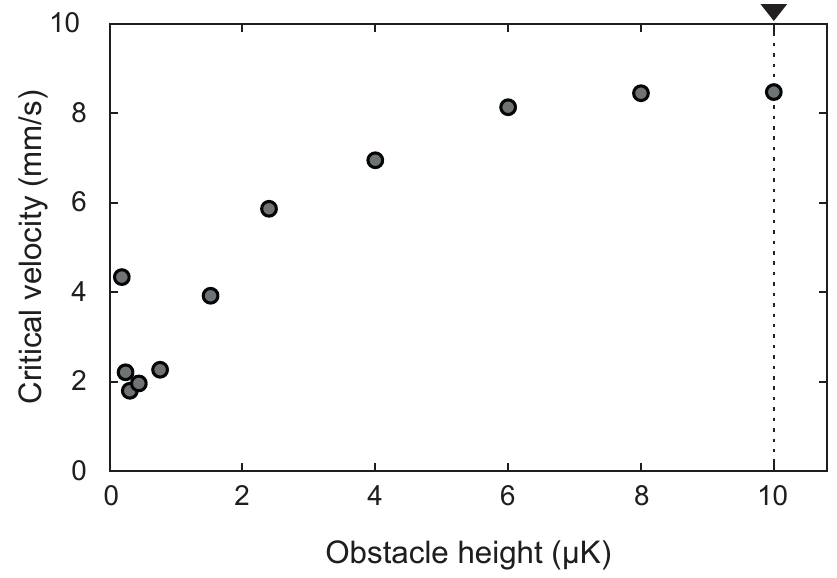}
\caption{Dependence of the vortex shedding critical velocity $v_{\rm c}$ on the obstacle height. The measurement was performed at unitarity, where $\mu \sim 0.6 E_{\rm F}$. The vertical dashed line marks the chosen obstacle height.}
\label{S2}
\end{figure}

Since the obstacle is a focused Gaussian laser beam, its soft boundary can affect the measured critical velocity~\cite{Kwon2015}. Our critical velocity measurements are performed in the hard-wall limit by setting the height of the obstacle about 20 times higher than the Fermi energy $E_{\rm F}$. The hard-wall limit is confirmed by examining the dependence of the critical velocity on the obstacle height [Fig.~\ref{S2}]. As described in Ref.~\cite{Kwon2015}, the critical velocity has a dip structure, reaching its minimum value when the obstacle height equals the chemical potential of the sample.

The waist of the Gaussian beam is assessed from \textit{in-situ} images of the sample. Accounting for the imaging resolution of our system ($\approx 5~\mu$m), a $1/e^2$ radius of $w_{0}=9.5 ~\mu$m is obtained. Hence, the obstacle is a macroscopic object in the condensate as it is an order of magnitude larger than the healing length and the pair size within the investigated interaction regimes. The obstacle diameter $D$ is computed from the ratio of the obstacle height to the chemical potential of the sample~\cite{Kwon2015}. For a Gaussian laser beam generating an obstacle potential $V(r) = V_{\rm 0}e^{-2r^2/w_{\rm 0}^2}$, the diameter of a density-depleted hole in a sample whose chemical potential is $\mu$ is given by 
\begin{align}
D = 2w_0\sqrt{{\rm ln}(V_0/\mu)/2}.
\label{Eq1}
\end{align}
Once the diameter is known for a given interaction strength, $D$ at other $1/k_{\rm F}a$'s can be obtained using $\mu$ from the equation of state~\cite{Manini2005}.

\subsection{Detecting vortices}

Since the radial confinement is dominantly provided by the residual magnetic curvature of the Feshbach field, we switch off both the Feshbach field and the ODT for time-of-flight expansion of the sample prior to taking an absorption image of the sample to detect vortices. During this process, fermionic pairs are converted into deeply bound molecules. After an expansion of 13.5 ms, the magnetic field is quickly ramped up to 690 G in 10 ms, where an absorption image is taken.

\begin{figure}[t]
\includegraphics[width=8.5cm]{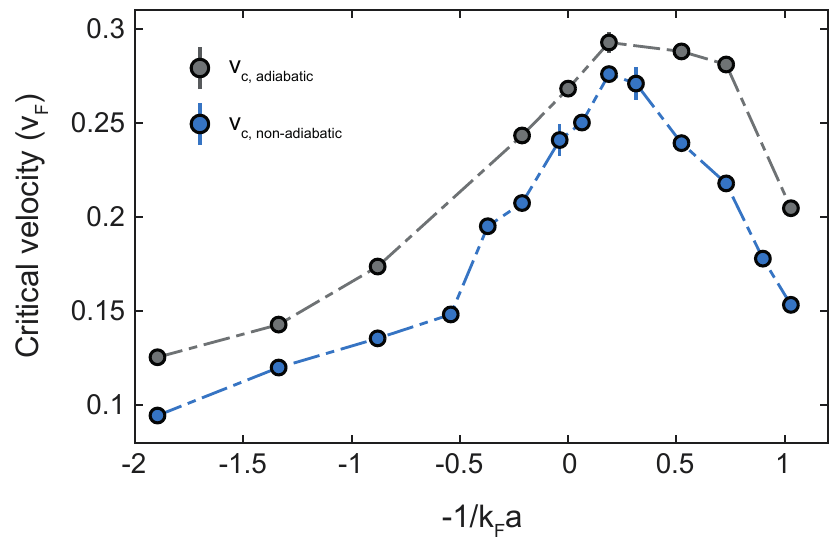}
\caption{Comparison of the vortex shedding critical velocity $v_{\rm c}$ between an adiabatic obstacle beam ramp down in 200 ms (gray) and a non-adiabatic quick ramp down in 30 ms (blue) as a function of $-1/k_{\rm F}a$.}
\label{S3}
\end{figure}

\subsection{Adiabatic switch-off of the obstacle beam}

\begin{figure}[t]
\includegraphics[width=7.5cm]{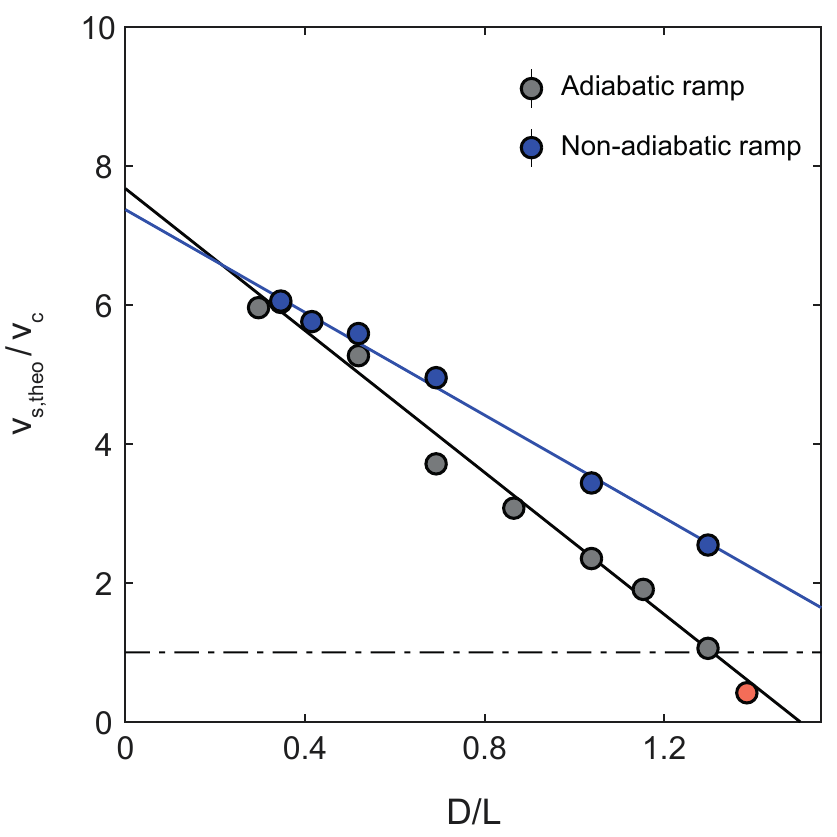}
\caption{Comparison of the vortex shedding critical velocity $v_{\rm c}$ between an adiabatic obstacle beam ramp down in 200 ms (gray) and a non-adiabatic quick ramp down in 30 ms (blue) as a function of $D/L$ at $-1/k_{\rm F}a = -1.9$.}
\label{S4}
\end{figure}

In the course of analyzing the adiabaticity of the obstacle beam intensity ramps, we found that a quicker ramp down reduces the measured critical velocity for vortex shedding. This observation is consistent with our description of the vortex shedding mechanism from energy considerations. A non-adiabatic switch off procedure of the obstacle beam will inject energy into the system, in addition to the energy provided by the drag force of the moving obstacle, lowering the measured critical velocity for vortex shedding. Since this argument holds for all interaction strengths, the decrease should be observed throughout the BEC-BCS crossover, and indeed this is experimentally confirmed, as shown in Fig.~\ref{S3}. Furthermore, the same argument can be made for the obstacle traveling distance $L$; a lower critical velocity should be measured when the obstacle beam ramp down is non-adiabatic. We verify this statement at 740 G $(-1/k_Fa=-1.90)$ where the critical velocity for vortex shedding is measured at various $L$ with the two different ramp down durations, 200 ms and 30 ms [Fig.~\ref{S4}].

\subsection{Measurement of the speed of sound}

For comparison with the measured vortex shedding critical velocities, the speed of sound is measured across the crossover. Initially, a weak density depletion is created at the center of the sample with the same obstacle laser beam that is used to measure the critical velocity. The beam is suddenly switched off, creating a circular density wave that propagates outwards. We measure the initial speed of the radial density wave propagation that equals the speed of sound in the center region of the fluid. The speed of sound measured this way corresponds to the speed of sound of the column averaged density since our sample is hydrodynamic in the $z$-direction~\cite{Weimer2015}.

\end{document}